\documentclass[aps,10pt,twocolumn,showpacs,prx,amsmath,amssymb,floatfix,superscriptaddress,dvipsnames]{revtex4-2}

\usepackage{amsthm}
\usepackage{mathtools}
\usepackage{physics}
\usepackage{bbm}
\usepackage{amsfonts}
\usepackage{amssymb}
\usepackage{graphicx}

\usepackage[T1]{fontenc}
\usepackage{bbold}
\usepackage{float}
\usepackage[utf8]{inputenc}
\usepackage{csquotes}

\usepackage[colorlinks=true,linktocpage=true]{hyperref}
\hypersetup{
    allcolors  = {blue},
}

\theoremstyle{plain}

\usepackage{array}
\usepackage{multirow}
\usepackage{xcolor}

\usepackage[utf8]{inputenc}
\usepackage[english]{babel}
\usepackage[T1]{fontenc}

\usepackage{amsfonts}
\usepackage{dsfont}
\usepackage{mathrsfs}
\usepackage{enumerate}
\usepackage{braket}
\usepackage{framed}
\usepackage{graphicx}

\usepackage{hyperref}
\usepackage{cleveref}
\usepackage{natbib}
\usepackage{csquotes}
\usepackage{xcolor}
\usepackage[normalem]{ulem}

\usepackage{bbold}

\newcommand{\bbar}{\mathchar'26\mkern-9mu b}

\newcommand{\luis}[1]{\textcolor{black}{{#1}}}
\newcommand{\ifusp}{Department of Mathematical Physics, Institute of Physics, University of São Paulo,\\ Rua do Matão 1371, São Paulo 05508-090, São Paulo, Brazil} 

\newcommand{\uoft}{Department of Chemistry and Centre for Quantum Information and Quantum Control,
University of Toronto, 80 Saint George St., Toronto, Ontario, M5S 3H6, Canada}

\usepackage{soul}

\begin{document}

\title{Canonical quantization for Equilibrium Thermodynamics}




\author{Luis F. Santos}
\email{luisf@usp.br}
\affiliation{\ifusp}

\author{Victor Hugo M. Ramos}
\email{vhmarques@usp.br}
\affiliation{\ifusp}

\author{Danilo Cius}
\email{cius@if.usp.br}
\affiliation{\ifusp}

\author{Mario C. Baldiotti}
\email{baldiotti@uel.br}
\affiliation{Departamento de Física, Universidade Estadual de Londrina, 86051-990, Londrina-PR, Brazil.}

\author{Bárbara Amaral}
\email{bamaral@if.usp.br}
\affiliation{\ifusp}
\affiliation{\uoft}
 
\date{\today}

\begin{abstract}
We formulate a canonical quantization of Equilibrium Thermodynamics by applying Dirac’s theory of constrained systems. Thermodynamic variables are treated as conjugate pairs of coordinates and momenta, allowing extensive and intensive quantities to be promoted to operators in a Hilbert space. The formalism is applied to the ideal gas, the van der Waals gas, and the photon gas, illustrating both first- and second-class quantization procedures. For the ideal gas, a Schrödinger-like equation emerges in which entropy plays the role of time, and the wave function acquires a phase determined by the internal energy. A pseudo-Hermitian framework restores Hermiticity of the temperature operator and establishes the equivalence among constraint realizations. The approach naturally leads to thermodynamic uncertainty relations and suggests extensions to quantum and topological phase transitions, as well as black-hole and non-equilibrium thermodynamics.

\end{abstract}

\maketitle
\section{Introduction}

The theory of Equilibrium Thermodynamics \cite{callen2006thermodynamics, blundell2010concepts, zemansky1998heat, buchdahl2009concepts} has been one of the most successful phenomenological frameworks in physics, remaining almost unchanged through successive scientific revolutions. Whether in Quantum Mechanics or General Relativity, in flat or curved spacetime, thermodynamics continues to provide valuable insights when applied in its traditional form. This robustness arises because thermodynamics disregards the internal structure of systems: all relevant physical quantities are defined in terms of variables accessible from outside the system under consideration. Consequently, the number of variables required to describe a system in thermodynamics is generally smaller than in other physical theories. For instance, Thermodynamic principles are employed to study both quantum and classical gases \cite{pathria2011statistical}, thermal machines \cite{PhysRevX.5.031044}, black holes \cite{hawking1976}, and even more exotic phenomena such as sonoluminescence \cite{young2004sonoluminescence} and heat transport between qubits \cite{naim}.

Despite its wide applicability, this generality comes at a price: being so different from other physical theories, Thermodynamics does not have a clear action principle formalism associated with it. The main physical theories, such as Electromagnetism, Quantum Field Theory, Quantum Mechanics, and General Relativity, can all be derived from an action whose extremization yields the equations of motion \cite{lemos2018analytical, goldstein1950classical}. 
In Thermodynamics, however, such a structure is far from obvious. Recently, several works have sought to construct an action-based formulation of Thermodynamics, employing Lagrangian \cite{Stokes_2017, BRAVETTI2015377, vaz2011lagrangian}, Hamiltonian \cite{peterson1979, baldiotti2016hamiltonian, ghosh2024hamiltonian}, and even Hamilton–Jacobi \cite{RAJEEV20082265} approaches. Although not as firmly established as in other fields, these descriptions reveal remarkable equivalences, particularly when viewed through the lens of Dirac’s formalism for constrained systems \cite{dirac2013lectures, sundermeyer1982constrained, lemos2018analytical, Prokhorov_Shabanov_2011, henneaux1992quantization}. In this framework, Thermodynamics can be clearly structured as a mechanical system with as many constraints as extensive variables, placing Analytical Mechanics and Equilibrium Thermodynamics on the same footing \cite{baldiotti2016hamiltonian}.

Building on this equivalence between Thermodynamics and Analytical Mechanics, the natural next step is to explore its quantum counterpart. Focusing on the Hamiltonian approach, in the present work we apply Dirac’s quantization procedure for constrained systems \cite{dirac2013lectures, henneaux1992quantization, gitman2012quantization, barcelos1987canonical, ita2023quantization, das2020lectures} to propose a direct quantization scheme for Equilibrium Thermodynamics. Although several previous works have attempted to quantize Thermodynamics directly \cite{RAJEEV2008768, FITZPATRICK20112384, GHOSH2022126402, maslov1994geometric, maslov2002quantized}, they generally do not follow Dirac’s procedure and fail to provide explicit descriptions of thermodynamic systems in the quantum regime. In particular, by neglecting the constrained nature of the theory, such approaches often lead to trivial Lagrangians (and Hamiltonians) that carry no physical content. In contrast, our formulation emphasizes the physically meaningful quantities and yields wave functions that can be analyzed within a well-known and physically consistent framework. \luis{We bypass statistics by taking only the classical equations of state, without invoking any thermodynamic limit.}

It is important to emphasize that our approach differs from quantization schemes based on contact manifolds \cite{RAJEEV2008768}. In fact, in this latter approach the resulting algebra of quantum observables has an intricate structure and uncertainty relations cannot be derived in a straightforward manner even for simple thermodynamic systems such as the ideal gas. Furthermore, the formalism presented here fundamentally differs from conventional quantum thermodynamics \cite{deffner2019quantum}. In the latter, one typically adopts a dynamical perspective, coupling a quantum system to a statistical-mechanical reservoir.  In contrast, our formulation treats the thermodynamic variables themselves as generalized coordinates and momenta, allowing a direct quantization of the thermodynamic system. As a result, quantities such as entropy, temperature, volume, and pressure acquire the status of operators acting in a Hilbert space. We expect that this procedure (and its consequences) will provide deeper insights into the nature of thermodynamic quantities, offering a more fundamental interpretation of what entropy and temperature are, why they emerge in a macroscopic description, and how they relate to the underlying mechanical structure. 

The perspective presented here, where thermodynamic variables are treated as operators, finds physical motivation in systems where quantum fluctuations, rather than thermal ones, become dominant. As argued in \cite{RAJEEV2008768}, we can consider, for example, a gas of cold atoms confined by a potential. Even if thermal fluctuations in aggregate quantities are small, the ``wall'' confining the gas has an inherent quantum uncertainty in its position, $\Delta x$, on the order of the atomic wavelength. This implies a volume uncertainty $\Delta V = A\Delta x$, where $A$ is the wall's area. Similarly, the pressure $P$, arising from momentum transfer at this wall, also becomes uncertain. Since the momentum uncertainty is $\Delta p \sim \hbar / \Delta x$, and the number of collisions per unit time is $\nu \sim \rho v A$ (where $\rho$ is the number density and $v$ is the typical velocity), the pressure uncertainty is $\Delta P \sim \hbar\rho v /\Delta x  $. This line of reasoning suggests a specific uncertainty principle for these thermodynamic variables, $\Delta P \Delta V \gtrsim \hbar_1$, where the effective Planck constant $\hbar_1 = \hbar \nu$ (with $\nu \approx \rho v A$) scales with the area $A$. \luis{This heuristic argument points toward a possible route for the direct quantization of macroscopic variables. However, when presented in this form, such relations remain system-specific and fail to exhibit the required extensivity, and therefore cannot be regarded as genuinely thermodynamic. In contrast, the quantization framework developed here yields general and extensive uncertainty relations that parallel those obtained through statistical mechanics and established approximation methods \cite{SCHLOGL1988679, BERNARDES1998186, TU1Uffink1999655, TU2publisi2017}.}


This work is organized as follows. In Section~\ref{Sec_2}, we present the general framework of classical constrained systems from a mechanical perspective, discuss their quantization, and show how Thermodynamics can be expressed in terms of a classical constrained system. This section serves as a brief review. In Section~\ref{section_3}, we combine these concepts and introduce an algorithm to quantize thermodynamic systems. Three illustrative examples are analyzed: the ideal gas, the van der Waals gas, and the photon gas. We place particular emphasis on the ideal gas, which serves as a comprehensive example of the possibilities offered by our procedure. In Section~\ref{sec4}, we discuss the physical consequences of the quantization, focusing mainly on the ideal gas. Topics include the equivalence between wave functions, entropy playing the role of a time parameter in Schrödinger-like equations, uncertainty principles, and probability currents. Finally, in Section~\ref{sec: conclusion}, we summarize our main results and outline future perspectives.






\section{Constrained Systems} \label{Sec_2}
\subsection{Classical systems}
\label{subsec_cosclass}

In classical mechanics \cite{lemos2018analytical, goldstein1950classical}, when a constraint appears in the description of a system, it is usually handled by explicitly eliminating the redundant variables before solving the dynamical equations. From a gauge theory perspective, this corresponds to fixing a specific gauge and removing all redundant degrees of freedom, thereby working with the minimal set of variables. While this procedure is convenient, it may obscure important information about the gauge structure of the system, particularly quantities that are gauge invariant and can provide deeper physical insight. To retain this information, we shall work in the extended phase space, where the constraints are treated as nonvanishing quantities, and study the consequences of this formulation. Naturally, the usual physical description must be recovered when our results are projected onto the surface where the constraints hold.

Let us consider a system described by $2N$ variables in the Hamiltonian formalism, subject to $M$ constraints defined as
\begin{equation}
    \phi_m (q,p) = 0,\quad m=0,1,...,M\leq 2N, \label{constraints}
\end{equation}
where $(q,p) \equiv (q_1,...,q_N, p_1,...,p_N)$ respectively denotes the generalized coordinates and their conjugate momenta. In this setting, the canonical Hamiltonian $\tilde{H}= \tilde{H}(q,p)$ is equivalent to the extended Hamiltonian \cite{dirac2013lectures, sundermeyer1982constrained, Prokhorov_Shabanov_2011}
\begin{equation}
    H=\tilde{H} + \sum_{m=1}^M \lambda ^{m} (t) \phi_m, \label{hamilton ext}
\end{equation}
where the functions $\lambda^{m} (t)$ are arbitrary multipliers that reflect the gauge freedom of the theory. This equivalence holds because the constraints satisfy $\phi_m = 0$ on-shell,  so both Hamiltonians generate the same dynamics. From now on, however, we will generally work off-shell, i.e., in the extended phase space where $\phi_m\neq 0$. 

From the definition of the Poisson bracket for two functions $f = f(q,p)$ and $g = g(q,p)$,
\begin{equation}
    \left\{ f,g\right\} =\sum_{i=1}^{N}\left(\frac{\partial f}{\partial q_{i}}\frac{\partial g}{\partial p_{i}}-\frac{\partial f}{\partial p_{i}}\frac{\partial g}{\partial q_{i}}\right),
\end{equation}
the Hamiltonian formalism provides the off-shell dynamical evolution of an observable $O = O(q,p)$ as
\begin{equation}
\frac{dO}{dt}=\{ O,\tilde{H}\} +\sum_{m=1}^{M}\lambda^{m}\left\{ O,\phi_{m}\right\} +\frac{\partial O}{\partial t}.
\end{equation}
We observe that all the redundancy of the theory is then encoded in the contributions $\lambda^m \{O, \phi_m\}$, showing that the Poisson brackets between the observable and the constraints play a central role. Indeed, if the set of constraints $\{\phi_m\}$ satisfies
\begin{equation}
    \left\{ \phi_{n},\phi_{m}\right\} =\sum_{j=1}^{M}f_{nm}^{j}\phi_{j}, \label{firstclass}
\end{equation}
where $f_{nm}^{j}$ are arbitrary functions of $(q,p)$, then two observables $O_{\lambda}(q,p)$ and $O_{\lambda'}(q,p)$, corresponding to different gauge choices $\lambda^m$ and $\lambda^{\prime\,m}$, differ in the first order by
\begin{align}
    \delta O&\equiv O_{\lambda}-O_{\lambda^\prime} \nonumber \\&\approx(t-t_{0})\sum_{m=1}^{M}(\lambda^{m}-\lambda^{\prime\, m})\left\{ O_{\lambda}\left(t_{0}\right),\phi_{m}\right\},
\end{align}
which means that the constraints $\phi_m$ act as generators of gauge symmetry transformations connecting different gauge orbits. Therefore, when the constraints satisfy condition (\ref{firstclass}), they are called \textit{first-class constraints}. If this condition is not fulfilled for some pair $\phi_\alpha$ and $\phi _\beta$, both constraints are said to be \textit{second-class}. As we shall see, systems possessing only first-class constraints admit a more straightforward quantization scheme. 

In the case of second-class constraints, one cannot associate them to a gauge symmetry. However, we can redefine the Poisson brackets in such a way that we can treat second-class constraints algebraically in the same way as first-class ones. These new brackets, called the \emph{Dirac brackets}, naturally remove the redundancy of the theory, making standard operations (such as quantization) clearer and more direct. They are defined, for a pair of functions $f(q,p)$ and $g(q,p)$, as
\begin{equation}
    \left\{ f,g\right\}_D\equiv\left\{ f,g\right\} -\sum_{\alpha,\beta }\left\{ f,\phi_{\alpha}\right\} \left(K^{-1}\right)^{\alpha\beta}\left\{ \phi_{\beta},g\right\}, \label{diracBK} 
\end{equation}
with
\begin{equation}
    K_{\alpha\beta}\equiv\left\{ \phi_{\alpha},\phi_{\beta}\right\}, \label{k matrix}
\end{equation} 
where the Greek indices label the second-class constraints. With this definition, we have $\left\{ \phi_{\alpha},f\right\}_D = 0$ for any function $f(q,p)$, which is equivalent to say that the constraints $\phi_{\alpha}$ do not generate any transformations on the dynamical quantities, as expected. In particular, if for some $\beta$, we choose $f = \phi_\beta$, we recover relation (\ref{firstclass}) by exchanging $\{ \cdot, \cdot\} \to \{ \cdot, \cdot\}_D$ and taking $f^{j}_{mn}=0$. If we now assume that Dirac brackets instead of Poisson brackets in the Hamiltonian formalism, one can show that all the redundancy in the theory is removed, and we can deal with the Hamiltonian system as an unconstrained one. In this scenario, one of the main challenges, of course, is computing the matrix $K_{\alpha\beta}$.

In the cases where only first-class constraints are present, we work with the usual Poisson brackets. More generally, we can decompose the problem into a subspace of first-class constraints (where we use Poisson brackets) and a subspace of second-class constraints (where we use Dirac brackets).

\subsection{Mechanical formulation of Thermodynamics}\label{sec: thermoconstrain}

We now turn our attention to Thermodynamics. In this section, we essentially summarize the main ideas presented in \cite{baldiotti2016hamiltonian}, where a Hamiltonian approach to Thermodynamics is developed. It is important to emphasize, however, that this construction is equivalent to the Hamilton–Jacobi formalism for Thermodynamics developed in \cite{RAJEEV20082265}. Both works formalize the central insight introduced in \cite{peterson1979}, where the author pointed out the strong analogy between Thermodynamics and Hamiltonian Classical Mechanics.

In Thermodynamics, the imposition of integrability conditions on the internal energy immediately leads to the Maxwell relations \cite{callen2006thermodynamics, zemansky1998heat}. Similarly, in Hamiltonian mechanics, imposing integrability conditions on generators of canonical transformations, yields Poisson brackets structure \cite{arnold1989mathematical}. Hence, identifying thermodynamic variables with coordinates and momenta of some phase space, thermodynamic integrability conditions can be rewritten as the canonical Poisson brackets. Both frameworks then share an underlying Hamiltonian structure. The main distinction, however, lies in the role of time: in Classical Mechanics, time plays a special role as an external parameter, whereas in Thermodynamics all variables stand on an equal footing. This difference makes it difficult to push the analogy further and often results in a seemingly trivial dynamics for Thermodynamics, with null Hamiltonians or Lagrangians that reduce to total derivatives. To overcome this issue, the approach proposed in \cite{baldiotti2016hamiltonian} and followed here is to regard Thermodynamics as a fully constrained mechanical system. In this way, the Hamiltonian may vanish on-shell, and yet remain nontrivial in the extended phase space [see Eq. (\ref{hamilton ext})].

To illustrate how this framework operates, let us briefly recall the structure of Thermodynamics. For a system described by its internal energy $U$, the first law can be written as
\begin{equation}
    dU = TdS + \sum_{i=1}^{N}Y^{i}dX_i, \label{first law}
\end{equation}
where $T$ is the temperature, $S$ the entropy, and $X_i$ ($Y^{i}$) represent the extensive (intensive) variables, such as volume and particle number (pressure and chemical potential), respectively. Then, one is usually interested in determining the internal energy $U$ as a function of the extensive variables, that is, finding $U = U(S, X)$.

To proceed, we first note that the number of degrees of freedom — that is, the number of independent exact differentials on the right-hand side of Eq. (\ref{first law}) — can always be reduced by one through the homogeneity of the theory. This simplification can be made explicit by introducing specific quantities, defined as $x_{\mu} \equiv X_{\mu}/\mathcal{N}$, where $\mathcal{N}$  represents one of the extensive quantities (usually number of particles in the system). In terms of these intensive variables, the first law takes the form
\begin{equation}
    du = Tds + \sum_{i=1}^{N-1}Y^idx_i. \label{firstlaw2}
\end{equation}
Hence, finding $u=u(s,x)$, with $x=(x_1,...,x_{N-1})$, characterizes a thermodynamic system with $N$ degrees of freedom. 

To accomplish this, one must specify $N$ equations of state of the form
\begin{equation}
    \phi_m (s, T, x, Y) = 0, \quad m = 1, 2, ..., N, \label{eq state}
\end{equation}
which correspond exactly to the number of independent degrees of freedom of the system. By solving these equations for all the intensive variables and substituting the results into Euler’s relation,
\begin{equation}
    u = Ts + \sum_{i=1}^{N-1} Y^{i}x_i + Y^{N}(s,x),
\end{equation}
we obtain the desired functional form $u = u(s, x)$, and all the equations of state follow from its partial derivatives. All equilibrium properties or relevant thermodynamic quantities can then be evaluated. 

Note, however, that Eq.~(\ref{eq state}) has exactly the same form as the constraints discussed earlier. This observation is highly suggestive: one may interpret Thermodynamics as a classical constrained system described by the tautological form given in Eq.~(\ref{firstlaw2}), with coordinates $(s, x) \equiv (\tau, q)$ and conjugate momenta $(T, Y) \equiv (\pi, p)$. The system is then fully constrained by the conditions $\phi_m(s, T, x, Y) = \phi_m(\tau, \pi, q, p) = 0$. This formulation eliminates the need to introduce an explicit time parameter, as there are no genuine dynamics in physical space. Nevertheless, we can extend the framework to an enlarged phase space, where $\phi_m (\tau, \pi, q, p) \neq 0$. In this extended space, the system is described by the Hamiltonian
\begin{equation}
    H = \sum_{m=1}^N\lambda^{m}\phi_m \label{hamiltonian}
\end{equation}
which generates the flow
\begin{equation}
    \boldsymbol{X}_{H}=\sum_{i=1}^{N-1}\left(\frac{\partial H}{\partial p^{i}}\frac{\partial}{\partial q_{i}}-\frac{\partial H}{\partial q^{i}}\frac{\partial}{\partial p_{i}}\right)+\frac{\partial H}{\partial\pi}\frac{\partial}{\partial\tau}-\frac{\partial H}{\partial\tau}\frac{\partial}{\partial\pi},
\end{equation}
allowing us to derive equations analogous to Hamilton’s equations of motion in this extended thermodynamic space. In particular, if the Hamiltonian takes the form
\begin{equation}
    H=\pi+h\left(p,q,\tau\right),
\end{equation}
then by the Hamilton formalism any thermodynamic observable $O_T$ evolves according to
\begin{equation}
    \frac{dO_T}{d\tau}=\frac{\partial O_T}{\partial \tau}+\left\{ O_T,H\right\} _{\left(q,p\right)}-\frac{\partial h}{\partial\tau}\frac{\partial O_T}{\partial\pi},
\end{equation}
where $\left\{ .,.\right\}_{\left(q,p\right)}$ denotes the Poisson brackets evaluated only over the variables ${q_i, p_i}$.

It is particularly interesting to note that in this formulation, the entropy $\tau$ plays the role of an evolution parameter. If $O_T$ does not depend explicitly on the temperature $\pi$, or if $h$ does not depend explicitly on the entropy $\tau$, one recovers Hamilton’s equations with entropy acting as the time variable. Applications of this formalism can be found in \cite{baldiotti2016hamiltonian}. In Section \ref{section_3}, however, we will provide some examples and extend this construction to include the quantization procedure. 

The mechanical formulation of Thermodynamics developed in this subsection shows that the thermodynamic variables $(s, x)$ and $(T, Y)$ naturally appear as conjugate pairs, analogous to $q_i$'s and $p_i$'s in Classical Mechanics. This observation makes it natural to ask whether Thermodynamics itself can be canonically quantized. Since it can be regarded as a fully constrained mechanical theory, we must employ a formalism capable of consistently handling such constraints. To this end, we conclude our review section by introducing the main concepts of Dirac’s quantization procedure for constrained systems. Combined with the formalism presented here, this framework will lead us to the desired quantization of Thermodynamics in the following section.

\subsection{Quantization of constrained systems}\label{sec: quantization}

Roughly speaking, the quantization of constrained systems follows naturally from canonical quantization \cite{dirac2013lectures, henneaux1992quantization}. In fact, if a system is described by first-class constraints, we promote the canonical coordinate pairs and constraints to operators, and the Poisson brackets to commutators in a suitable Hilbert space, as in the standard quantization scheme. The operator algebra is then automatically preserved, and the dynamically relevant states are obtained by imposing that the constraint operators annihilate them. In the case of second-class constraints, however, we must first replace the Poisson brackets with Dirac brackets quantization, and then promote the modified brackets to commutators. This typically leads to nonstandard commutation relations among the canonical variables, requiring us to work consistently with this new algebra when defining the dynamics and promoting observables to operators. In what follows, we present the general framework for both first- and second-class constraints.

\subsubsection{First-class constraints\label{sec: firstclass}}

Let us again consider a classical system subject to $M$ constraints, as defined in Eq. (\ref{constraints}), all of which satisfy the first-class condition given by Eq. (\ref{firstclass}). Using the standard Poisson brackets, one can proceed with the usual canonical quantization procedure \cite{dirac1981principles, shankar2012principles}:
\begin{equation}
    \left\{ q_{i},p_{j}\right\} =\delta_{ij}\rightarrow\left[\hat{q}_{i},\hat{p}_{j}\right]=i\hbar\delta_{ij},
\end{equation}
where $\delta_{ij}$ is the Kronecker delta and $\hbar$ is Dirac's constant. The hats indicate operators acting on a suitable Hilbert space $\mathscr{H}$, which is called the kinematical Hilbert space. Since all classical observables are functions of $(q,p)$, this correspondence allows us to construct the corresponding quantum observables directly, up to the usual operator-ordering ambiguities \cite{ferialdi2023}. In particular, the constraint functions are promoted to operators via the substitution $\phi_m (q,p) \rightarrow \hat{\phi}_m (\hat{q}, \hat{p})$.

To obtain the states $\ket{\psi} \in \mathscr{H}$ compatible with the constraint structure, one imposes the conditions
\begin{equation}\label{eq:DynamicalHilbertSpace}
    \hat{\phi}_{i}\left| \psi\right\rangle =0, \quad 1\leq i \leq M,
\end{equation}
which defines the dynamical subspace $\mathscr{H}_D \subset \mathscr{H}$. By this definition, we can write \begin{align}
    \mathscr{H}_D = \bigcap_{j=1}^M \ker(\hat{\phi}_j),
\end{align} which is in fact a linear subspace of $\mathscr{H}$. Within this subspace, the constraint operators satisfy
\begin{equation}
    [\hat{\phi}_{i},\hat{\phi}_{j}]=0,
\end{equation}
as a direct consequence of the first-class condition and the quantization procedure. Hence, in $\mathscr{H}_D$ the algebra of constraints closes consistently, ensuring the internal compatibility of the quantization procedure. Furthermore, we denote the remaining portion of the Hilbert space as $\mathscr{H}_R = \mathscr{H}\setminus\mathscr{H}_D$.

 In the coordinate representation, where the operators $\hat{q}_i$ act multiplicatively,
\begin{equation}
    \hat{q}_i\ket{\psi} = q_i\ket{\psi}, \quad q_i \in \mathbb{R}
\end{equation}
the conjugate momenta \luis{take} the standard form,
\begin{equation}
    \hat{p}_i\left|\psi\right\rangle =-i\hbar\frac{\partial }{\partial q_i}\ket{\psi}.
\end{equation}
In this representation, the constraint conditions can be translated into differential equations for the wave functions $\psi (q) \equiv \left\langle q_1,...,q_N\right.\left|\psi\right\rangle $, i.e.,
\begin{equation}
    \hat{\phi}_m(\psi) \equiv \bra{q_1,...,q_N}\hat{\phi}_m\ket{\psi} = 0. \label{condition}
\end{equation}
If the system is fully constrained (i.e., $M=2N$) and all constraints are first-class, Eq.~(\ref{condition}) can determine the explicit form of the wave function $\psi(q)$. For instance, in the case of a nonrelativistic free particle with redundant degrees of freedom, Eq.~(\ref{condition}) reproduces the standard Schrödinger equation \cite{henneaux1992quantization}. 

\subsubsection{Second-class constraints}\label{sec: secclass}

If the system contains second-class constraints, that is,
\begin{equation}
 \left\{ \phi_{\alpha},\phi_{\beta}\right\} \neq \sum_{\gamma=1}^{M}f_{\alpha\beta}^{\gamma}\phi_{\gamma},
\end{equation}
we can no longer promote the constraints $\phi_{\alpha}$ directly to operators and impose the condition (\ref{eq:DynamicalHilbertSpace}) to determine the relevant dynamical states $\ket{\psi}$. Indeed, this would imply
\begin{equation}
    [\hat{\phi}_{\alpha},\hat{\phi}_{\beta}] \ket{\psi }=0 \Rightarrow [\hat{\phi}_{\alpha},\hat{\phi}_{\beta}]=0
\end{equation}
in $\mathscr{H}_D$. This is contradictory, since the nonvanishing Poisson brackets of the classical constraints imply that the corresponding second-class constraint operators do not, in general, commute in $\mathscr{H}_D$.

To overcome this difficulty, we promote Dirac’s brackets (Eq. \ref{diracBK}) to commutators in $\mathscr{H}$, since they preserve the same algebraic structure as Poisson brackets and, consequently, as quantum commutators \cite{das2020lectures, barcelos1987canonical, ita2023quantization}. In this framework,
\begin{equation}
    \left\{ q_{i},p_{j}\right\}_D =F_{ij}(q,p)\rightarrow\left[\hat{q}_{i},\hat{p}_{j}\right]=i\hbar\hat{F}_{ij}(\hat{q},\hat{p}),
\end{equation}
where $F_{ij}(q,p)$ is given by the right-hand side of the Dirac bracket definition in Eq.~(\ref{diracBK}). It is important to stress that, in this case, the constraint conditions cannot be imposed to determine the physical states or wave functions as in Eq.~(\ref{condition}), since $\hat{\phi}_{\alpha}$ are not generators of gauge transformations. Instead, one must explicitly solve the Dirac brackets to remove the redundancies among the operators and thereby obtain the dynamical Hilbert space. Without a corresponding dynamical equation, however, it is generally difficult to derive explicit wave equations as in the first-class constraints case.

Having summarized the quantization of Hamiltonian systems presenting first- and second-class constraints, we will now proceed to apply this procedure to Thermodynamics.

\section{Dirac's Quantization for Thermodynamics}
\label{section_3}

Up to this point, we have established how to quantize constrained mechanical systems and how to describe Thermodynamics in terms of this framework. Next, we connect these two ideas. In broad terms, our approach follows the algorithm below:

\begin{enumerate}
\item \textbf{Mapping:} Identify the correspondence between thermodynamic quantities and the canonical variables (coordinates and momenta). This serves as a dictionary that allows us to express the thermodynamic system in a familiar mechanical form.
\item \textbf{Constraints:} Use the equations of state to identify a corresponding set of constraints
\begin{equation}
   \phi_\mu(\tau,\pi, q, p) = 0, 
\end{equation}
defined in the extended phase space. In particular, whenever possible, it is advantageous to cast the constraints in the form
\begin{equation}
    \phi_{\mu} = p_{\mu} + h_{\mu}(p,q,\tau), \label{mario}
\end{equation}
 as this structure typically ensures that they satisfy the first-class condition. Here, $p_{0}=\pi$, and each function $h_{\mu}(p,q,\tau)$ encapsulates the thermodynamic relations specific to the system under consideration.
\item \textbf{Classification:} Determine whether these constraints are of first or second class.
\item \textbf{First-class case:} If all the constraints are of first class, quantize them so that they become Hermitian operators in the kinematical Hilbert space $\mathscr{H}$. Then impose the condition given by Eq.~(\ref{condition}) to obtain the wave functions that describe the system.
\item \textbf{Second-class case:} If there are second-class constraints, solve the commutation algebra among the operators to eliminate redundancies and specify the dynamical Hilbert space $\mathscr{H}_D$.
\end{enumerate}

The general procedure outlined above can be best understood through explicit examples. In what follows, we present concrete realizations of quantized thermodynamic systems, illustrating each step of the algorithm. These examples form the core of our main results and demonstrate how the canonical quantization scheme can be consistently applied to Thermodynamics.

\subsection{Ideal gas} \label{subsection_idealgas}

Consider a system composed of $\mathcal{N}$ non-interacting particles confined in a volume $V$. Since the number of particles is conserved, it is convenient to work with specific quantities defined as $s \equiv S/\mathcal{N}$ and $x_i \equiv X_i/\mathcal{N}$. The equations of state for this system are given by \cite{zemansky1998heat}
\begin{align}
    P\left(T,v\right)=\frac{k_{B}T}{v} \quad \text{and} \quad T\left(u,v\right)=\frac{2}{3k_{B}}u, \label{stateGASIDEIAL} 
\end{align}
where $P$ denotes the pressure of the gas. The first law [Eq.~(\ref{firstlaw2})] for this system can thus be written as
\begin{equation}
    du = Tds - Pdv,
\end{equation}
which allows us to establish the correspondence $\left(\tau,\pi\right)=\left(s,T\right)$ and $\left(q,p\right)=\left(v,-P\right)$. With this mapping, the equations of state (\ref{stateGASIDEIAL}) become
\begin{align}
    \pi=-\frac{pq}{k_{B}} \quad \text{and} \quad 
    \pi=\frac{2}{3k_{B}}u.
\end{align}

The first constraint arises naturally from the first equation of state,
\begin{equation}
    \phi_{1}=\pi+\frac{pq}{k_{B}}. \label{phi1 gas ideal}
\end{equation}
Since $du$ represents the tautological form, we must eliminate the function $u(\tau,q)$ from the second equation of state to define a second constraint that depends only on the coordinates and momenta.
Solving for $u(\tau, q)$, we obtain
\begin{equation}
    u(\tau, q)=\frac{3}{2}Ae^{\frac{2\tau}{3k_{B}}}q^{-2/3}, \quad A\in\mathbb{R}^{+}, \label{u pro gas ideal}
\end{equation}
which leads to the second constraint,
\begin{equation}
    \phi_{2}= p+Ae^{\frac{2\tau}{3k_{B}}}q^{-5/3}. \label{phi2 gas ideal} 
\end{equation}

Since we are dealing with a two-dimensional thermodynamic system, two is precisely the number of constraints required to characterize it fully. We can therefore proceed with our algorithm. First, it is straightforward to verify that
\begin{equation}
    \left\{ \phi_{1},\phi_{2}\right\}  = \frac{\phi_{2}}{k_{B}}, \label{pp gas ideal}
\end{equation}
which means that both $\phi_1$ and $\phi_2$ are first-class constraints. 

Now that we have a complete set of first-class constraints, we can quantize the system directly by promoting the variables $\tau$, $\pi$, $q$ and $p$ to operators acting suitable linear subspaces of the Hilbert space $\mathscr{H} = L^2([\tau_{\text{min}},\tau_{\text{max}}]\times[q_{\text{min}},q_{\text{max}}],d\tau dq)$, where $q_{\text{max}}$ represents the maximum volume attainable by the gas container without rupture, $q_{\text{min}}$ its minimum volume, $\tau_{\text{max}}$ the maximum entropy, and $\tau_{\text{min}}$ the minimum entropy (which \luis{must be a non-negative constant} due to the Nernst theorem \cite{martin-olalla2025nernst}). In principle, one could consider $\tau_{\text{max}} \rightarrow \infty $, but this would render $\hat{\pi}$ (on the minimal domain $C^\infty_0([\tau_{\textrm{min}},\infty))$) not self-adjoint, and with no self-adjoint extensions \cite{gitman2012selfadjoint,Bonneau2001, BALDIOTTI20111630}. To avoid this \luis{technical} issue, we initially adopt a finite $\tau_{\text{max}}$. In fact, we promote the Poisson brackets
\begin{equation}
    \left\{ q,p\right\} =\left\{ \tau,\pi\right\} =1
\end{equation}
(all others vanishing) to commutators,
\begin{equation}
    \left[\hat{\tau},\hat{\pi}\right]	=\left[\hat{q},\hat{p}\right]=i\bbar. \label{commutadors ideal gas}
\end{equation}
In the expression above, 
$\bbar$ is a real, positive constant with dimensions of energy. \luis{As discussed in Sec. \ref{uncertainty}, we expect that $\bbar$ encodes the quantum nature of Thermodynamics.}

We can now promote the constraints to operators acting on suitable linear subspaces of $\mathscr{H}$. As a matter of fact, the second constraint $\phi_2$ involves no ordering ambiguity because $\left[\hat{q},\hat{\tau}\right]= 0$. Therefore, it can be promoted directly by
\begin{equation}
    \phi_{2}\rightarrow\hat{\phi}_{2}=\hat{p}-Ae^{\frac{2\hat{\tau}}{3k_{B}}}\hat{q}^{-5/3}.
\end{equation}
On the other hand, since $\left[\hat{q},\hat{p}\right]= i\bbar$, there are several possible operator orderings for $\hat{\phi}_1$ that reproduce the correct classical limit. To ensure that $\hat{\phi}_1$ is formally Hermitian, we adopt a symmetrized ordering for the $pq$ term:
\begin{equation}
    \phi_{1}\rightarrow\hat{\phi}_{1}=\hat{\pi}+\frac{\hat{p}\hat{q}+\hat{q}\hat{p}}{2k_{B}} = \hat{\pi}+\frac{\hat{q}\hat{p}}{k_{B}}-\frac{i\bbar}{2k_{B}}, \label{phi1simetrico}
\end{equation}
where the last equality follows from the commutation relation in Eq.~(\ref{commutadors ideal gas}).
As we will see, this specific choice of operator ordering has important consequences. For now, note that it leads to the correct algebra between the constraint operators: 
\begin{equation}
    [\hat{\phi}_{1},\hat{\phi}_{2}]=\frac{i \bbar}{k_{B}}\hat{\phi}_{2}.
\end{equation}

Having constructed the constraint operators in a consistent manner, we can now turn to the physical states. We focus on wave functions of the form $\psi(\tau, q)$, which represent the system in the coordinate representation. For this to be possible, we realize the algebra given in Eq.~(\ref{commutadors ideal gas}) as
\begin{subequations}
\begin{align}
    \hat{q}&\left|\psi\right\rangle =q\left|\psi\right\rangle, \label{q} \\
 \hat{p}&\left|\psi\right\rangle =-i\bbar\frac{\partial}{\partial q}\ket{\psi}, \label{p} \\ \hat{\tau}&\left|\psi\right\rangle =\tau\left|\psi\right\rangle, \label{tau} \\
 \hat{\pi}&\left|\psi\right\rangle =-i\bbar\frac{\partial}{\partial\tau}\ket{\psi}. \label{pi}
\end{align}
\end{subequations}
These realizations satisfy the commutation relations of Eq.~(\ref{commutadors ideal gas}) and allow us to express the constraint equations as differential equations for the wave function.


To obtain the physical states, we now apply the constraint conditions given in Eq.~(\ref{condition}), which for this case become
\begin{subequations}
\begin{align}
    \hat{\phi}_{1}(\psi)&=\left(\hat{\pi}+\frac{\hat{q}\hat{p}}{k_{B}}-\frac{i\bbar}{2k_{B}}\right)\psi\left(\tau,q\right)=0, \label{eq1 ideal}
    \\
    \hat{\phi}_{2}(\psi)&=\left(-i\bbar\frac{\partial}{\partial q}+Ae^{\frac{2\tau}{3k_{B}}}q^{-5/3}\right)\psi\left(\tau,q\right)=0. \label{eq2 ideal}
\end{align}
\end{subequations}
The only function that satisfies both equations simultaneously is
\begin{equation}\label{eq:WaveFunctionSimmetrycOrderdering}
    \psi\left(\tau, q\right)=\alpha e^{-\frac{\tau}{2k_{B}}}\exp\left[-\frac{3}{2i\bbar}Ae^{\frac{2\tau}{3k_{B}}}q^{-2/3}\right],
\end{equation}
where $\alpha$ is a normalization constant. Recalling the internal energy of the ideal gas, Eq.~(\ref{u pro gas ideal}), we can express the \textit{thermodynamic wave function} of an ideal gas as
\begin{equation}
    \psi\left(\tau, q\right)=\alpha e^{-\frac{\tau}{2k_{B}}}\exp\left[\frac{i}{\bbar}u\left(\tau, q\right)\right]. \label{psi gas ideal}
\end{equation}
 The similarity between this wave function and those governed by Hamiltonians in standard Quantum Mechanics is remarkable, a point we will explore shortly. It is also worth noticing that a shift in the internal energy, $ u\left(\tau, q\right) \rightarrow u\left(\tau, q\right) + \text{const.} $, changes the wave function only by an overall phase factor. From a quantum-mechanical perspective, this represents a symmetry transformation. A function of the same form was presented in \cite{isidro2018contact}, though derived there purely by analogy, without invoking Dirac’s quantization framework as performed here.   

 To ensure that the thermodynamic wave function possesses the same probabilistic interpretation as a conventional quantum-mechanical one, we must impose normalization to unity. For this purpose, we adopt the standard norm induced by the inner product in $\mathscr{H}$. This guarantees that the expectation values of operators correspond to physical quantities in the usual probabilistic sense.

Imposing the normalization condition $\left|\left|\psi\right|\right|^{2}\stackrel{!}{=}1$, we obtain 
 \begin{align}
      \left|\alpha\right|^{2}
     =\frac{e^{\frac{\tau_{\text{max}}+\tau_{\text{min}}}{2k_{B}}}}{2k_{B}\left(q_{\text{max}}-q_{\text{min}}\right)\sinh{\left(\frac{\tau_{\text{max}}-\tau_{\text{min}}}{2k_{B}}\right)}}.
 \end{align}

    With the normalization established, we can now analyze how the operators act on the thermodynamic wave function. Since $\hat{\tau}$ and $\hat{q}$ are multiplicative operators, it is straightforward to verify that they are both formally Hermitian and their expectation values yield the corresponding classical quantities. For the pressure operator $\hat{p}$, we have
    \begin{equation}
        \hat{p}\left(\psi\right)=-i\bbar\frac{\partial\psi}{\partial q}=\frac{\partial u}{\partial q}\psi,
    \end{equation}
    which confirms that $\hat{p}$ is also formally Hermitian and corresponds to $\partial_qu$, as in the classical case. However, for the temperature operator $\hat{\pi}$, we obtain
    \begin{equation}
    \label{eq:eigenequation_temperatureOp}
        \hat{\pi}\left(\psi\right)=-i\bbar\frac{\partial\psi}{\partial\tau} = \left(\frac{i\bbar}{2k_{B}}+\frac{\partial u}{\partial\tau}\right)\psi,
    \end{equation}
    which reproduces the classical result up to an additional imaginary shift. This extra term originates from the symmetrization of $\hat{\phi}_1$ performed earlier to ensure its formal Hermiticity [Eq. (\ref{phi1simetrico})]. As a consequence, the temperature acquires this small complex correction, which can be interpreted as a purely quantum feature with no classical analog. In the $\bbar \rightarrow 0$ limit, this shift vanishes, and we recover the standard thermodynamic relation.
    
    At first glance, the presence of the imaginary term in the temperature operator may seem to compromise the Hermiticity of $\hat{\phi}_1$. However, this feature is precisely what ensures its Hermitian character. Within the domain defined by the single wave function $\psi(\tau, q)$ obtained above, the operator combination $pq +qp$ is itself non-Hermitian, and its non-Hermitian contribution exactly cancels that of $\hat{\pi}$. The detailed demonstration of this cancellation is presented in Appendix~\ref{Appendix_A}.
    
    We will examine the physical implications of this imaginary shift in temperature in a later section. Before doing so, let us present the case of the van der Waals gas and show that all results obtained there are equivalent to those of the ideal gas. This will justify continuing our analysis using the ideal gas as a representative example, without any loss of generality.
    
\subsection{van der Waals gas}

Let us now turn our attention to a more realistic model: the van der Waals gas. Its equations of state are
\begin{subequations}
\begin{align}
    T\left(u,v\right)&=\frac{2}{3\luis{k_B}}\left(u+\frac{a}{v}\right), \label{vander1}
    \\ 
    P\left(T,v\right)&=\frac{T\luis{k_B}}{v-w}-\frac{a}{v^{2}}, \label{vander2}
\end{align}
\end{subequations}
where the parameters $w$ and $a$ account for the finite size of the particles and the attractive interactions between them, respectively. The van der Waals gas thus provides a more accurate description of real gases, particularly because it captures phase-transition behavior absent in the ideal-gas model \cite{callen2006thermodynamics, zemansky1998heat}.

Using the same mapping adopted for the ideal gas, $\left(s,T,v,P\right)=\left(\tau,\pi,q,-p\right)$,we can rewrite Eqs.~(\ref{vander1}) and (\ref{vander2}) as
\begin{subequations}
\begin{align}
    \pi&=\frac{2}{3k_{B}}\left(u+\frac{a}{q}\right), \label{pi vander} 
    \\
    p&=\frac{a}{q^{2}}-\frac{\pi k_{B}}{q-w}. \label{pvander}
\end{align}
\end{subequations}
As in the ideal gas case, Eq.~(\ref{pvander}) for $p$ immediately provides a constraint, which we define as
\begin{equation}
    \phi_{1}=\pi+\frac{\left(q-w\right)}{k_{B}}\left(p-\frac{a}{q^{2}}\right),
\end{equation}
again having the general form $\pi + h(q,p)$. This structure will naturally lead to entropy $\tau$ acting as a time-like parameter.

To obtain the second constraint, we use the expression for the internal energy,
\begin{equation}
    u_W=Ae^{\frac{2\tau}{3k_{B}}}\left(q-w\right)^{-\frac{2}{3}}-\frac{a}{q}, \label{energy vander}
\end{equation}
where the subscript $W$ is used to distinguish this internal energy from that of the ideal gas. Differentiating Eq.~(\ref{energy vander}) with respect to $q$, we obtain the pressure, which allows us to define the second constraint as
\begin{equation}
    \phi_{2}=p-\frac{a}{q^{2}}+\frac{2}{3}\left(q-w\right)^{-\frac{5}{3}}Ae^{\frac{2\tau}{3k_{B}}}.
\end{equation}
Once again, the first-class condition is satisfied,
\begin{equation}
    \left\{ \phi_{1},\phi_{2}\right\} = \frac{\phi_{2}}{k_{B}},
\end{equation}
and the quantization procedure follows directly, in complete analogy with the ideal gas case.

Without any ambiguity, we can promote the second constraint to an operator as
\begin{equation}
    \phi_{2}\rightarrow\hat{\phi}_{2}=\hat{p}-a\hat{q}^{-2}+\frac{2}{3}\left(\hat{q}-w\right)^{-\frac{5}{3}}Ae^{\frac{2\hat{\tau}}{3k_{B}}}.
\end{equation}
For $\phi_{1}$, as before, we must symmetrize to ensure Hermiticity, thus $\phi_1\rightarrow\hat{\phi}_1$ gives
\begin{align}
    \hat{\phi}_{1} = \hat{\pi}+\frac{1}{k_{B}}\left(\hat{q}\hat{p}-\frac{i\bbar}{2}-a\hat{q}^{-1}-w\hat{p}+wa\hat{q}^{-2}\right).
\end{align}
Considering the commutation relations in Eq.~(\ref{commutadors ideal gas}), we find
\begin{equation}
[\hat{\phi}_{1},\hat{\phi}_{2}]=i\bbar\frac{\hat{\phi}_{2}}{k_{B}},
\end{equation}
showing that the algebra of constraints is preserved, and all necessary ingredients are in place to construct the corresponding physical states.

Imposing the constraint conditions given in Eq.~(\ref{condition}) and using the explicit form of canonical operators (\ref{q}-\ref{pi}), we obtain
\begin{widetext}
\begin{subequations}
\begin{align}
    \hat{\phi}_{1}\left(\psi\left(\tau,q\right)\right)
    &= 
    \left[-i\bbar\frac{\partial}{\partial\tau}-\frac{i\bbar\left(q-w\right)}{k_{B}}\frac{\partial}{\partial q}-\frac{i\bbar}{2k_{B}}-\frac{a}{k_{B}}\frac{1}{q}+\frac{wa}{k_{B}}\frac{1}{q^{2}}\right]\psi\left(\tau,q\right)=0,
    \\
    \hat{\phi}_{2}\left(\psi\left(\tau,q\right)\right)
    &= 
    \left[-ib\frac{\partial}{\partial q}-\frac{a}{q^{2}}+\frac{2}{3}\left(q-w\right)^{-\frac{5}{3}}Ae^{\frac{2\tau}{3k_{B}}}\right]\psi\left(\tau,q\right)=0.
\end{align}
\end{subequations}
\end{widetext}
Both equations are simultaneously satisfied by the thermodynamic wave function
\begin{equation}
    \psi_W\left(\tau,q\right)=\alpha_{W}e^{-\frac{\tau}{2k_{B}}}\exp\left[\frac{i}{\bbar}u_W\left(\tau,q\right)\right],
\end{equation}
where $\alpha_W$ is the normalization constant.

Remarkably, this function has exactly the same form as that of the ideal gas, Eq.~(\ref{psi gas ideal}). Consequently, all results obtained previously, including the appearance of the imaginary contribution to the temperature, also hold here.
This structural similarity is consistent with the canonical transformation that relates the ideal gas and van der Waals gas, as discussed in \cite{baldiotti2016hamiltonian}. It also significantly extends the discussion presented in \cite{alarcon2018van, coutant2008quantum}.

With the structural equivalence between the ideal gas and the van der Waals gas now clearly established, we will henceforth use the ideal gas as our reference model when discussing the physical consequences of thermodynamic quantization (Sec. \ref{sec4}), unless otherwise stated. This choice simplifies the analysis without any loss of generality, since both systems share the same quantum structure under the Dirac quantization scheme.

\subsection{Photon gas and second-class constraints}

We conclude the presentation of the quantization procedure for thermodynamic systems by analyzing both first-class and second-class scenarios within a single model: the photon gas \cite{Leff2002, Leff2015}. In this system, the gas consists of noninteracting photons that can be emitted or absorbed by the walls of the container. Although the number of particles is not conserved, the vanishing chemical potential allows us to work with specific (per-particle) thermodynamic quantities. We proceed in two stages. First, we use the equations of state directly to define constraints of the form (\ref{mario}), showing that this leads to a first-class constrained system whose quantization is analogous to the previously addressed systems. Then, by introducing an additional redundancy into the description (through constraints that are not in the form of Eq.~(\ref{mario})), we obtain a second-class constrained model. In this case, the quantization procedure determines the dynamical Hilbert space but does not yield an explicit thermodynamic wave function. 

\subsubsection{First-class approach}

One way to express the equations of state for the photon gas is
\begin{subequations}
\begin{align}
    s=\frac{4\sigma}{3} vT^3 \quad \text{and} \quad   
    P=\frac{\sigma}{3}T^4, \label{photon2}
\end{align}
\end{subequations}
where 
\begin{equation}
    \sigma =\frac{8\pi^{5}k_{B}^{4}}{15c^{3}\hbar^{3}},
\end{equation}
is the three-dimensional Stefan-Boltzmann constant \cite{Baldiotti_2023}. Aiming to construct constraints of the form (\ref{mario}), we can manipulate these equations to express the internal energy as
\begin{equation}
    u_P (s,v)= K\frac{s^{4/3}}{v^{1/3}}+u_0, \label{uPHOTON}
\end{equation}
where $K$ and $u_0$ are real constants. From this expression, we can evaluate the temperature and pressure as
\begin{subequations}
    \begin{align}
 &T=\frac{\partial u_P}{\partial s}=\frac{4K}{3}\frac{s^{1/3}}{v^{1/3}}, \\
 &P=-\frac{\partial u_P}{\partial v}=\frac{K}{3}\frac{s^{4/3}}{v^{4/3}}.
    \end{align}
\end{subequations}

Using now the correspondence $(s,T,v,P)=(\tau,\pi,q,-p)$, we can then define the constraints as
\begin{subequations}
    \begin{align}
 &\phi_{1}\equiv\pi-\frac{4K}{3}\frac{\tau^{1/3}}{q^{1/3}}, \\
 &\phi_{2}\equiv-p-\frac{K}{3}\frac{\tau^{4/3}}{q^{4/3}}.
    \end{align}
\end{subequations}
It follows immediately that
\begin{equation}
    \left\{ \phi_{1},\phi_{2}\right\}  = 0,
\end{equation}
showing that both constraints are of first class. We can therefore proceed with a direct quantization by imposing condition~(\ref{condition}) and selecting a suitable realization for the canonical operator algebra. 

Using the standard representation given by Eqs.~(\ref{q}–\ref{pi}), we obtain that the thermodynamic wave function $\psi_P$ must satisfy the pair of equations
\begin{subequations}
    \begin{align}
            &\left(i\bbar \frac{\partial}{\partial\tau}+\frac{4K}{3}\frac{\tau^{1/3}}{q^{1/3}}\right)\psi_P=0,\\
            &\left(i\bbar \frac{\partial}{\partial q}-\frac{K}{3}\frac{\tau^{4/3}}{q^{4/3}}\right)\psi_P=0.
    \end{align}
    \end{subequations}
    Solving these two conditions simultaneously leads to
    \begin{equation}
        \psi_P = \alpha_P \exp\left[\frac{i}{\bbar}u_P(\tau,q)\right],
    \end{equation}
where $\alpha_P$ is the normalization constant and $u_P(\tau,q)$ is given by (\ref{uPHOTON}) via correspondence $(s,v)=(\tau,q)$.

This expression closely resembles the results obtained for the ideal and van der Waals gases, particularly when the constraint $\phi_1$ is not symmetrized [Eq.~(\ref{exemploNOsime})]. The structural similarity among these cases reinforces the internal consistency of the quantization framework and suggests that the ideal gas can serve as a representative model for exploring the physical implications and interpretative aspects of thermodynamic quantization in greater detail (Sec.~\ref{sec4}). 

Before turning to these interpretative aspects, however, it is instructive to examine how a second-class formulation can also arise within thermodynamic systems. To illustrate this, we continue our analysis of the photon gas, now introducing a distinct set of constraints that lead to a genuinely second-class structure.

\subsubsection{Second-class approach}

  A second-class formulation for the photon gas requires discarding constraints of the form (\ref{mario}). In fact, since the equations of state can always be cast in that form, to obtain second-class constraints, we shall not assume them as our starting point. Instead, we fix one thermodynamic variable and study the system's behavior on the corresponding hypersurface. In this spirit, we consider the photon gas under isentropic conditions, i.e., with constant entropy. On this surface, the thermodynamic behavior is given by the adiabatic relation
    \begin{align} Pv^{4/3}&=\xi,\quad\xi\in\mathbb{R}^{+}. \label{adiabatica}
    \end{align} 
    Using the dictionary $\left(s,T,v,P\right)=\left(\tau,\pi,q,-p\right)$, the adiabatic relation combined with the second expression in (\ref{photon2}) enables us to define the constraints
\begin{subequations}
    \begin{align}
      \phi_1 &= p+\frac{\sigma}{3}\pi^{4}, \\
      \phi_2 &= \xi q^{-4/3}+p.
    \end{align}
\end{subequations}
The corresponding Poisson brackets reads
    \begin{align}
        \left\{ \phi_{1},\phi_{2}\right\} &=\frac{4}{3}\xi q^{-7/3}, 
    \end{align}
which shows that $\phi_1$ and $\phi_2$ are of second class. Therefore, the system cannot be quantized as in the ideal or van der Waals cases, since these constraints do not correspond to gauge symmetries. In this case, one must instead construct the Dirac brackets [Eq.~(\ref{diracBK})] and promote them directly to commutators in the quantum formalism. 

The $\textbf{K}$ matrix, Eq. (\ref{k matrix}), for this case is
\begin{equation}
\boldsymbol{K}=
\left(
    \begin{array}{cc}
     0 & \frac{4}{3}\xi q^{-7/3}\\
     -\frac{4}{3}\xi q^{-7/3} & 0
    \end{array}
\right),
\end{equation}
which leads to
\begin{equation}
\boldsymbol{K}^{-1}=
\left(
    \begin{array}{cc}
    0 & -\frac{3}{4\xi}q^{7/3}\\
    \frac{3}{4\xi}q^{7/3} & 0
    \end{array}
\right).
\end{equation}
Thus, the Dirac brackets for two arbitrary functions $f = f(\tau, \pi, q, p)$ and $g= g(\tau, \pi, q, p)$ are
\begin{align}
    \left\{ f,g\right\} _{D}=\left\{ f,g\right\} 
    &+\frac{3q^{7/3}}{4\xi}\left\{ f,\phi_{1}\right\} \left\{ \phi_{2},g\right\} 
    \nonumber\\
    &-\frac{3q^{7/3}}{4\xi}\left\{ f,\phi_{2}\right\} \left\{ \phi_{1},g\right\} .
\end{align}

Next, we evaluate the Dirac brackets of the canonical variables to set up the quantization. The nonvanishing ones are
\begin{subequations}
\begin{align}
\left\{ \tau,\pi\right\} _{D}	&=1,
\\ \left\{ \tau,q\right\} _{D}	&=-\frac{\sigma}{\xi}\pi^{3}q^{7/3},
\\ \left\{ \tau,p\right\} _{D}	&=\frac{4}{3}\sigma\pi^{3}.
\end{align}
\end{subequations}
Following the quantization procedure, we promote these Dirac brackets to commutators in a Hilbert space, obtaining
\begin{subequations}
\begin{align}
    \left[\hat{\tau},\hat{\pi}\right]	&=i\bbar,
    \\ 
    \left[\hat{\tau},\hat{q}\right]	&=-i\bbar\frac{\sigma}{\xi}\hat{\pi}^{3}\luis{\hat{q}}^{7/3},
    \\ 
    \left[\hat{\tau},\hat{p}\right]	&=-i\bbar\frac{4}{3}\sigma\hat{\pi}^{3}.
\end{align}
\end{subequations}
The only way to satisfy this algebra is by choosing
\begin{subequations}
\begin{align}
\hat{q}	&=\left(\frac{\sigma}{3\xi}\hat{\pi}^{4}+C\right)^{-3/4}, \label{qPHOTON} \\
\hat{p}	&=-\frac{\sigma}{3}\hat{\pi}^{4}, \label{pPHOTON}
\end{align}
\end{subequations}
where $C$ is a constant. With these definitions, all redundancies generated by $\phi_1$ and $\phi_2$ are eliminated, allowing us to define the dynamical Hilbert space of the system. In this scheme, the physical states are described by square-integrable functions of the form $\psi_P = \psi_P(\tau)$. Whenever we need to evaluate expectation values or analyze the behavior of the pressure and volume, their corresponding operator representations given by Eqs.~(\ref{qPHOTON}) and~(\ref{pPHOTON}) must be used. Importantly, the fact that the wave function depends only in $\tau$ is not accidental: each value of the entropy labels a distinct isentropic surface, and within this approach, the quantum description is constructed on such fixed-entropy hypersurface of the thermodynamic space. 

This second-class formulation is presented solely to illustrate the consistency of the Dirac–bracket approach when thermodynamic constraints do not generate gauge symmetries. Interestingly, a closely related structure appears in the ADM formulation of General Relativity using the Palatini action \cite{Montesinos2020}. In that analogy, space-time foliation plays a role similar to selecting isentropic surfaces, while treating the metric and affine connection as an independent fields mirrors the introduction of the adiabatic relation (\ref{adiabatica}) as an additional equation of state.

Having established how the procedure works in this example, we will not pursue its physical implications further. Instead, we now turn our attention to the ideal gas, which provides a clearer and more insightful setting for analyzing the physical consequences of thermodynamic quantization.

\section{Physical implications for the ideal gas} \label{sec4}

\luis{For the examples discussed in the previous section, it is important to emphasize that the quantization procedure presented here differs fundamentally from the standard statistical-mechanical approach to quantum gases. In the latter, one begins with the conventional quantum-mechanical Hamiltonian describing a collection of microscopic particles and then employs, typically, partition functions to derive the thermodynamic equations of state in the thermodynamic limit \cite{pathria2011statistical}.}

\luis{In contrast, our approach proceeds in the opposite direction: we directly quantize thermodynamic quantities themselves, promoting them to operators. In this sense, the system continues to be treated as a “black box”, characterized only by macroscopic variables; however, this box is now described by a wave function. Although abstract, this perspective allows us to reinterpret equilibrium thermodynamics from a novel standpoint, opening the door to new conceptual insights at a foundational level. In what follows, we use the quantized ideal gas model to explore some of the implications and insights afforded by this framework.}

\subsection{Schrödinger-like equation and probability flow}

Having established the quantum formulation of the ideal gas, we can now examine some of its direct consequences. The physical condition (\ref{condition}) can be interpreted, in the spirit of a Hamiltonian constraint (\ref{hamiltonian}), as
\begin{equation}
    \hat{H}\left(\psi\right)=\hat{\phi}_{1}\left(\psi\right)+\kappa\hat{\phi}_{2}\left(\psi\right)=0,\quad \kappa \equiv  \frac{\lambda^{1}}{\lambda^{2}}.
\end{equation}
Using the explicit form of $\hat{\phi}_1$, this relation can be rewritten as
\begin{equation}
i\bbar\frac{\partial\psi\left(\tau,q\right)}{\partial\tau}=\hat{h}\left(\hat{q},\hat{p}\right)\psi\left(\tau,q\right), \label{schrodinger}
\end{equation}
where we employed $\hat{\phi}_2 (\psi)=0$ and defined
\begin{equation}
    \hat{h}\left(\hat{q},\hat{p}\right) \equiv \frac{\hat{q}\hat{p}}{k_{B}}-\frac{i\bbar}{2k_{B}}. 
\end{equation}
Eq. (\ref{schrodinger}) is thus a Schrödinger-like equation, with entropy $\tau$ playing the role of a time parameter and $\hat{h}\left(\hat{q},\hat{p}\right)$ acting as a non-Hermitian Hamiltonian. The wave function $\psi(\tau, q)$ therefore describes a non-conservative evolution under entropic translation, an intuitively consistent picture, since entropy changes are expected to accompany temperature variations (recall that $\hat{h}=-\hat{\pi}$ represents the temperature operator). 

\luis{Moreover, since the operator $\hat{h}$ is non-Hermitian, the emergent “evolution” of this closed system is inherently non-unitary. This feature stands in marked contrast to the usual formulations of Quantum Thermodynamics, yet it is by no means contradictory. Indeed, for an isolated thermodynamic system, the second law of Thermodynamics dictates that entropy can only increase. This intrinsic irreversibility establishes a preferred direction for natural processes, which is commonly associated with the arrow of time \cite{Parrondo_2009}.}

\luis{At first sight, choosing entropy as a time parameter may appear to be merely a convenient option. In principle, within the extended phase-space formulation, one may select any thermodynamic quantity to play the role of a “time parameter”. However, entropy is distinguished by a unique property: by virtue of the second law, it is the only thermodynamic quantity that is necessarily monotonic \cite{buchdahl2009concepts}. This makes it not just a possible choice, but the most natural and consistent one.}

\luis{Moreover, it is well known that temperature and entropy cannot be measured directly \cite{zemansky1998heat, buchdahl2009concepts}. All instruments designed to infer these quantities operate through the measurement of mechanical variables, such as volume or pressure. In this sense, temperature is always obtained indirectly. Therefore, the fact that temperature is not an observable in the strict quantum–mechanical sense should not be surprising. What must be avoided, however, is the possibility that genuinely mechanical quantities become non-Hermitian.}

\luis{Within our formalism, adopting entropy as the evolution parameter leads, in the case of the ideal gas, to a naturally emergent non-unitary dynamics. This feature encodes the irreversibility intrinsic to thermodynamic processes. In this way, the framework provides further support for the well-known hypothesis connecting entropy with temporal irreversibility.}



One may further define the probability of finding the system in any volume within the interval $[q_{\text{min}}, q_{\text{max}}]$ as a function of entropy $\tau$ as
\begin{equation}
    \mathcal{P}(\tau)  \equiv \int_{q_{\text{min}}}^{q_{\text{max}}} \psi^*(\tau,q)\psi(\tau,q)dq,  
\end{equation}
and we can manipulate Eq. (\ref{schrodinger}) to show that
\begin{align}
    \frac{d\mathcal{P}(\tau)}{d\tau}=\int_{q_{\text{min}}}^{q_{\text{max}}}\frac{\psi^{*}\hat{h}\left(\psi\right)-\psi\hat{h}^{\dagger}\left(\psi^{*}\right)}{i\bbar}dq=-\frac{2e^{-\frac{\tau}{k_{B}}}}{k_B}.
\end{align}
This result shows that the system stabilizes as $\tau\rightarrow \infty $. In this limit, the non-Hermitian contribution (associated with the complex part of the temperature) is exponentially suppressed by entropy growth. Consequently, entropy increase can be interpreted as a mechanism driving the emergence of classicality within this thermodynamic quantum framework.  

It is important to emphasize that the approach developed in this section does not violate Pauli’s theorem on the quantization of the time parameter \cite{galapon2002pauli}. The reason is that all quantization performed here takes place within an extended phase space. This situation is analogous to the treatment of time in reparametrization-invariant theories, where one introduces an extended space to describe, for instance, the nonrelativistic free particle within the constrained Hamiltonian formalism. In this case, time is initially promoted to a canonical coordinate, but once the system is projected onto the dynamical subspace, it resumes its role as an evolution parameter, and the standard Schrödinger equation is recovered \cite{henneaux1992quantization}.

\subsection{Restoring Hermiticity: Pseudo-Hermitian Approach and $\hat{\phi}_1$ Equivalence} \label{subsec_pseudo}

Another way to deal with the imaginary temperature is considering an \emph{entropy-dependent pseudo-Hermitian approach} which is directly analogous to the well-established time-dependent pseudo-Hermitian approach (see Appendix \ref{Appendix_C}) 
\cite{mostafazadeh2010, fring2016, fring2017, kesley2021}. 

The motivation for this approach is subtle. In the Schrödinger-like equation, Eq. \eqref{schrodinger}, entropy $\tau$ acts as the evolution parameter while $\hat{\pi}$ up to a sign plays the role of the Hamiltonian. The operator $\hat{\pi}$ is manifestly non-Hermitian due to a constant imaginary term. Critically, its complex eigenvalues, shown in Eq. \eqref{eq:eigenequation_temperatureOp}, confirm it cannot be a physical observable. Furthermore, it is not pseudo-Hermitian, meaning it cannot be mapped to a Hermitian operator via an entropy-independent similarity transformation.

However, the formal analogy between this entropy-driven evolution and a time-dependent quantum system allows us to employ the time-dependent pseudo-Hermitian formalism. By treating $-\hat{\pi}$ as the non-observable generator of evolution and $\tau$ as the time parameter, this framework provides the relations needed to construct an associated Hermitian operator $\hat{\varpi}$. 

To find the explicit form of $\hat{\varpi}$, we introduce an invertible Dyson map $\hat{\eta}(\tau)$ and define a new state $\chi(\tau,q)=\hat{\eta}(\tau)\psi(\tau,q)$. The new state $\chi$ must obey a new Schrödinger equation
\begin{equation}
i\bbar\frac{\partial\chi\left(\tau,q\right)}{\partial\tau}=\hat{\varpi}\left(\hat{q},\hat{p}\right)\chi\left(\tau,q\right), \label{eq:schrodinger_chi}
\end{equation}
where the new generator $\hat{\varpi}$ is related to the original generator ($-\hat{\pi}$) by the transformation 
\begin{equation}
\label{eq:hermitian_temperatureOp}
\hat{\varpi} = \hat{\eta}(\tau)(-\hat{\pi})\hat{\eta}^{-1}(\tau) + i\bbar\left(\frac{\partial\hat{\eta}(\tau)}{\partial\tau}\right)\hat{\eta}^{-1}(\tau).
\end{equation}
By choosing the appropriate Dyson map, 
\begin{equation}
\hat{\eta}(\tau)=e^{\frac{\tau}{2k_{B}}},    
\end{equation}
that precisely cancels the anti-Hermitian part of $-\hat{\pi}$, the calculation successfully identifies the physical temperature as the Hermitian operator 
\begin{equation}
\hat{\varpi} = \frac{\hat{q}\hat{p}}{k_{B}}.
\end{equation}
This $\hat{\varpi}$ represents the physical temperature observable in the Hermitian representation (i.e., the $\chi$-space). 

Nevertheless, since our description is done in the original $\psi$ representation, we must find the corresponding pseudo-Hermitian operator $\hat{\Pi}$ that acts in this representation. The operator $\hat{\Pi}$ is related to $\hat{\varpi}$ via the inverse similarity transformation 
\begin{equation}
\hat{\Pi} = \hat{\eta}^{-1}(\tau)\hat{\varpi}\hat{\eta}(\tau).
\end{equation}
In this specific case, the Dyson map $\hat{\eta}(\tau)$ is a function of $\tau$, which is assumed to commute with $\hat{q}$ and $\hat{p}$. This leads to the simple but important result $\hat{\Pi} = \hat{\varpi}$. We can see this explicitly by rearranging the Eq. \eqref{eq:hermitian_temperatureOp} into the form $\hat{\Pi} = -\hat{\pi} +i\bbar\hat{\eta}^{-1}(\tau)\frac{\partial\hat{\eta}(\tau)}{\partial\tau}$. 
Substituting the chosen Dyson map confirms the result 
\begin{equation} 
\hat{\Pi} = -\hat{\pi} + \frac{i\bbar}{2k_{B}} = \frac{\hat{q}\hat{p}}{k_{B}}. 
\end{equation}
Therefore, the operator $\hat{q}\hat{p}/k_{B}$ represents the physical temperature in both representations. In the $\chi$-space, it is Hermitian. In the $\psi$-space, it is pseudo-Hermitian with respect to the modified inner product
\begin{equation}
\langle\cdot,\cdot\rangle_{\Theta(\tau)}:=\langle\cdot,\hat{\Theta}(\tau)\cdot\rangle,
\end{equation}
with 
$\hat{\Theta}(\tau)=\hat{\eta}^{\dagger}(\tau)\hat{\eta}(\tau)=e^{\frac{\tau}{k_{B}}}$ 
being the entropy-dependent metric operator \cite{mostafazadeh2013pseudo}. Note that this holds true for all other relevant observables of the theory. Thus, we can establish the set of thermodynamic observables $\{\hat{q},\hat{p},\hat{\tau},\hat{\Pi}\}$, all of which are pseudo-Hermitian with respect to the new inner product. 

A key insight from this pseudo-Hermitian treatment is that the resulting wave function, $\chi(\tau,q)$, is equivalent to the one that would have been obtained by choosing a simple, asymmetric operator ordering from the start (such as those discussed in Appendix \ref{Appendix_B}) 
\begin{equation}
    \phi_1 \rightarrow \hat{\phi}_{1}=\hat{\pi}+\frac{\hat{q}\hat{p}}{k_{B}},
\end{equation}
i.e., an ordering that does not introduce an imaginary term. We can show a similar equality for another choice of $\hat{p}\hat{q}$ ordering, which allows us to conclude that this formalism maps all constrained operator-ordering choices to an equivalent physical description. Thus, for the ideal gas, one can therefore either start with the symmetrized (non-Hermitian) $\hat{\pi}$ and use the entropy-dependent pseudo-Hermitian formalism to find the physical operator $\hat{\Pi}$, or start with an asymmetric (and already Hermitian) operator from the beginning. The fact that both paths lead to equivalent dynamics elucidates the deep consistency of the formalism. 

\subsection{\luis{Uncertainty relations and $\bbar$ constant}}\label{uncertainty}

When thermodynamic quantities are promoted to operators in a Hilbert space according to the commutation rules of Eq.~(\ref{commutadors ideal gas}), it is natural to ask whether uncertainty relations also arise. For any pair of Hermitian operators $(\hat{A},\hat{B})$ satisfying $[\hat{A},\hat{B}]=i\gamma$ with $\gamma \in \mathbb{R}$, the uncertainty principle holds, i.e.,
\begin{equation}
    \Delta \hat{A}\Delta \hat{B}\geq\frac{\gamma}{2},
\end{equation}
where 
$\Delta \hat{O}\equiv\sqrt{\langle \hat{O}^{2}\rangle -\langle\hat{O}\rangle^{2}}$ is the standard deviation. Note the calculation of the expectation value $\langle \hat{O} \rangle$ depends on the metric employed, although the final physical value must be the same (see Eq.~\eqref{eq:app_expectationvalue}). It is crucial that the choice of state and the choice of operator realization are consistent. Physical quantities must be calculated using the operators and states from the same representation: either the Hermitian operators (e.g., $\hat{\varpi}$) with the $\ket{\chi}$ states, or the pseudo-Hermitian operators (e.g., $\hat{\Pi}$) with the $\ket{\psi}$ states and the $\hat{\Theta}(\tau)$ metric.

In the thermodynamic case, we have shown that even though the temperature operator $\hat{\pi}$ is not strictly Hermitian, it can be associated with a pseudo-Hermitian counterpart preserving the same commutation algebra. This ensures that the algebraic structure of the theory remains intact, allowing us to treat the extensive and intensive thermodynamic variables as conjugate pairs of coordinates and momenta within a mechanical framework. From Eq.~(\ref{commutadors ideal gas}), we immediately obtain the fundamental thermodynamic uncertainty relations
\begin{subequations}
\begin{align}
    \Delta s\Delta T &\geq\frac{\bbar}{2}, 
    \\
    \Delta v\Delta P &\geq\frac{\bbar}{2},
\end{align}
\end{subequations}
which we interpret as \textit{Equilibrium Thermodynamic Uncertainty Relations}. It is important to contrast the physical scales at which these uncertainty relations apply with those discussed in the introduction, i.e. $\Delta P \Delta V \gtrsim \hbar_1$. The latter is system-dependent, scaling with the boundary area $A$. In contrast, the relations derived here depend on the constant $\bbar$. This follows from the genuinely thermodynamic perspective we employ: the systems are regarded as black boxes, with their internal boundaries and microscopic details left completely unspecified.

Moreover, if we choose to express the first law in its entropic form \cite{callen2006thermodynamics},
\begin{equation}
    ds=\frac{du}{T}+\frac{P}{T}dv,
\end{equation}
we may promote the quantities $u$, $\beta_1\equiv T^{-1}$, $v$ and $\beta_2\equiv PT^{-1}$ to operators acting on the Hilbert space\luis{, satisfying the canonical commutation relations}
\begin{equation}
    \luis{\left[\hat{u},\hat{\beta}_1\right]	=\left[\hat{v},\hat{\beta}_2\right]=i\bbar_0 \label{commutadorsENTROPY},}
\end{equation}
\luis{where $\bbar_0$ is a real, positive constant with units of $k_B$.} 

\luis{In this representation, it follows directly that}
\begin{subequations}
\begin{align}
    \Delta u \Delta \left(\frac{1}{T}\right)&\geq\frac{\bbar_0}{2},  \label{incertezaTOP1}
    \\
    \Delta v \Delta \left(\frac{P}{T}\right) &\geq\frac{\bbar_0}{2}. \label{incertezaTOP2}
\end{align}
\end{subequations}
\luis{It is worth emphasizing that previous results on equilibrium thermodynamic uncertainty relations were derived using statistical and approximate methods \cite{SCHLOGL1988679, BERNARDES1998186, TU1Uffink1999655, TU2publisi2017}. In particular, within the entropic formulation, one finds}
    \begin{subequations}
\begin{align}
    \tilde{\Delta} u \tilde{\Delta} \left(\frac{1}{T}\right)&\geq k_B, \label{incerCLASS1}
    \\
    \tilde{\Delta} v \tilde{\Delta} \left(\frac{P}{T}\right) &\geq k_B \label{incerCLASS2}
\end{align}
\end{subequations}
\luis{$\tilde{\Delta}$ indicates the standard statistical dispersion, rather than quantum uncertainties defined through operator expectation values.}

\luis{Remarkably, in \cite{miller2018energy} it is shown that the dispersion $\tilde{\Delta}$ is purely classical, and that the bound becomes larger when quantum fluctuations are taken into account. This indicates that Eqs.~(\ref{incerCLASS1}–\ref{incerCLASS2}) arise solely from thermal fluctuations, reflecting our statistical ignorance about the system.}

\luis{In contrast, our formalism is fully operatorial. Therefore, Eqs.~(\ref{incertezaTOP1}–\ref{incertezaTOP2}) are not expected to contain any classical uncertainty in the sense of statistical ignorance. Instead, our relations should be interpreted as refined bounds on the dispersion that capture only the genuinely quantum contribution (namely, the part that is independent of statistical ignorance). Since quantum fluctuations are expected to be smaller than classical statistical ones, we assume that $\bbar_0/2 \leq k_B$.}


\luis{Considering that we are dealing with a fully constrained theory, thermodynamic systems (as illustrated in the previous examples) are described by a single wave function, i.e., thermodynamic states are pure states. One therefore expects that, at a fundamental level, only this genuinely quantum uncertainty is present, while any classical contribution should be understood as arising from experimental limitations rather than from the theory itself.}

\luis{Accordingly, one may expect that violations of Eqs.~(\ref{incerCLASS1}–\ref{incerCLASS2}) could in principle be observed, thereby allowing an estimate of $\bbar_0$. In the spirit of \cite{miller2018energy}, one can expect that quantum correlations will increase the lower bound of $\Delta u \Delta (1/T)$. Thus, in first order in $\bbar_0$, we can schematically represent the full (classical plus quantum) uncertainty relation as}
\begin{equation}
    \luis{\Delta u \Delta \left(\frac{1}{k_B T}\right) 	\gtrsim  1+ \frac{\kappa}{\Delta u^2}\bbar_0},
\end{equation}
\luis{where $\kappa$ encodes information about the quantum state of the system. The first term on the right-hand side of the above equation corresponds to the classical (statistical) contribution [Eq.~(\ref{incerCLASS1})], while the second term, governed by $\kappa$, represents the genuinely quantum contribution. For pure states, one expects the classical contribution to be absent, so that violations of Eq.~(\ref{incerCLASS1}) may occur. Because the energy fluctuation $\Delta u$ is directly proportional to the heat capacity $c_v$ \cite{miller2018energy}, the limit $\Delta u \rightarrow 0$ is naturally reached as $c_v \rightarrow 0$. According to the third law of thermodynamics \cite{martin-olalla2025nernst}, this condition is satisfied in the low-temperature regime ($T \rightarrow 0$). In this limit, the quantum term dominates ($\kappa \Delta u^{-2}\bbar_0 \gg 1$), characterizing a purely quantum thermodynamic state free from classical statistical uncertainties. By quantifying the classical bound violation, one can estimate the magnitude order of $\bbar_0$.}

\luis{This reasoning may also be used to estimate the magnitude order of $\bbar$, since consistency between the energetic and entropic formulations of Thermodynamics requires that both yield results at the same characteristic scale. In particular, their corresponding “quantum” versions should be governed by parameters of the same order of magnitude.}

\luis{Is worth emphasizing that this approach addresses an emblematic question originally raised by Bohr and Heisenberg: whether intrinsic uncertainty relations between thermodynamic quantities should exist even at equilibrium, analogous to those in quantum mechanics \cite{TU1Uffink1999655}. Previous attempts to resolve this issue have remained ambiguous, largely due to the absence of an operatorial structure in equilibrium Thermodynamics. The formalism developed here provides precisely such a structure, thereby offering a natural framework in which these thermodynamic uncertainty relations emerge.} 

\luis{Finally, it is important to stress that the uncertainty relations derived in this framework should not be interpreted as statistical fluctuations in the usual sense. In standard statistical mechanics, fluctuations arise from microscopic degrees of freedom and are associated with ensemble averages and thermal agitation. By contrast, in the present approach the uncertainties are intrinsic and originate from the quantum nature of the macroscopic thermodynamic variables themselves. Since quantities such as temperature, entropy, pressure, and volume are promoted to quantum operators, they cannot, in general, be simultaneously specified with arbitrary precision. The resulting uncertainty relations therefore reflect the fact that the macroscopic system is described by a thermodynamic wave function, rather than by sharply defined state variables. While for mechanical quantities like volume and pressure one may heuristically relate this behavior to collective coordinates and momenta of the constituent particles, this analogy is not essential. Fundamentally, the uncertainties arise because the thermodynamic state is treated as a quantum object, and its defining variables naturally fluctuate within finite intervals, even in the absence of statistical or thermal noise.}
\section{Summary and outlook}\label{sec: conclusion}
In this work, we have reinterpreted Equilibrium Thermodynamics as a classical constrained mechanical system, and subsequently applied Dirac’s quantization procedure to formulate its quantum counterpart. Through this approach, we developed a consistent quantum theory of Thermodynamics, in which the extensive and intensive variables (such as entropy, temperature, volume, and pressure) are promoted to operators acting in a Hilbert space.

With this framework, unlike earlier approaches found in the literature, we were able to explicitly obtain wave functions for both the ideal gas and the van der Waals gas, as well as to characterize the physical Hilbert space associated with the photon gas. We observed that whenever the constrained Hamiltonian takes the form $H = \pi + h(q,p)$, the system naturally admits a Schrödinger-like equation, in which entropy plays the role of a time-like parameter. Furthermore, the resulting thermodynamic wave function exhibits a phase governed by the internal energy, closely analogous to the time-evolution operator in standard Quantum Mechanics. 

We also discussed several physical consequences of this quantization procedure, focusing on the ideal gas as a representative case. In this context, we found that the evolution generated by the constraints is intrinsically non-unitary when expressed in terms of entropy as the evolution parameter. This naturally aligns with the thermodynamic arrow of time: entropy increases monotonically, selecting a preferred direction for physical processes and embedding irreversibility directly into the quantum description. Moreover, we observed that the growth of entropy naturally leads to the emergence of classicality, as the non-Hermitian components of the theory become exponentially suppressed in the high-entropy limit. We further derived a set of thermodynamic uncertainty relations, establishing direct analogies with the canonical uncertainty principles of Quantum Mechanics. Additionally, we demonstrated that different realizations of the constraints (particularly the various representations of $\hat{\phi}_1$) are physically equivalent within a pseudo-Hermitian framework, which also provides a consistent way to interpret the imaginary component of the temperature operator as a physically meaningful, gauge-related feature of the formalism.

For future work, we plan to explore further physical consequences of this formalism, with particular emphasis on what can be extracted from the wave functions themselves. In particular, we aim to investigate how quantum phase transitions (especially topological phase transitions \cite{ventuti2007}) can be incorporated into this framework. Moreover, our approach opens the door to the quantization of more complex thermodynamic systems, such as black holes \cite{BALDIOTTI201722, Dereli_2019} and Bose gases described by a quasi-thermodynamic approach \cite{Baldiotti_2023}. 

\section*{Acknowledgments}

LFS acknowledges the financial support of São Paulo Research Foundation (FAPESP) under grant number 2024/08114-0. VHMR acknowledges the financial support of Coordenação de Aperfeiçoamento de Pessoal de Nível Superior (CAPES) – Brazil, Finance Code 001, and gratefully acknowledges the kind hospitality of the Department of Mathematics of the University of Genova. BA acknowledges financial support from the Instituto Serrapilheira, Chamada No. 4 2020. and the  Fundação de Amparo à Pesquisa do Estado de São Paulo, Auxílio à Pesquisa—Jovem Pesquisador, through Grant No. 2020/06454-7 (until September 2025). 

\bibliography{references}

\appendix
\onecolumngrid

\section{Hermiticity of $\hat{\phi}_1$ for ideal gas}
\label{Appendix_A}
In this appendix we discuss the hermiticity of the operator $\hat{\phi}_1$ with the symmetric ordering, which represents one of the constraints of the canonically quantized ideal gas. This operator, defined in Eq. (\ref{phi1simetrico}), is given by $\hat{\phi}_1 = \hat{\pi} + \hat{A}$, where $\hat{\pi}$ is the canonical pair of $\hat{\tau}$, and $\hat{A} = (\hat{p}\hat{q}+\hat{q}\hat{p})/2k_B$. 

\subsection{Operator $\hat{A} = \frac{\hat{p}\hat{q}+\hat{q}\hat{p}}{2 k_{B}}$}

Let us consider the operator $\hat{A}$ with its domain in the dynamical Hilbert space, i.e., the one characterized by states $|\psi\rangle$ such that $\phi_{m}|\psi\rangle=0$, where $\phi_{m}$ are the constraints. Thus, noticing that we have only one state as the solution to the constraints, we obtain
\begin{align}
\hat{A}(\psi) = -\dfrac{ib}{2 k_B}\left( \psi(\tau,q) +2q\frac{\partial\psi}{\partial q} \right).
\end{align}
Using the explicit solution for $\psi(\tau,q)$ from Eq.~\eqref{eq:WaveFunctionSimmetrycOrderdering}, we see that $\psi(\tau,q)\overline{\psi(\tau,q)}$ does not depend on $q$. Hence, we obtain
\begin{equation}
\left\langle \psi \right| \hat{A} \left| \psi \right\rangle - \left\langle \psi \right| \hat{A}^\dagger  \left| \psi \right\rangle = \dfrac{i\bbar}{k_B},
\end{equation}
and therefore, the operator $A$, in the domain of the dynamical Hilbert space, is not an hermitian operator.

\subsection{Operators $\hat{\pi}$ and $\hat{\phi}_{1}$}

Analogously, we will analyze the hermiticity of the operator $\hat{\pi}$, whose action on a state in the dynamical Hilbert space is given by 
\begin{equation}
\hat{\pi}(\psi) = \left(\frac{ib}{2k_{B}} + \frac{\partial u}{\partial\tau}\right)\psi(\tau,q) .
\end{equation}
Hence, the operator $\hat{\pi}$ is not hermitian because
\begin{align*}
\left\langle \psi \right| \hat{\pi} \left| \psi \right\rangle - \left\langle \psi \right| \hat{\pi}^\dagger  \left| \psi \right\rangle  =- \dfrac{i\bbar}{k_B}.
\end{align*}
However, as the operator $\hat{\phi}_1 = \hat{\pi}+\hat{A}$, the non-hermitian contributions cancel out, i.e. \begin{align}
    \left\langle \psi \right| \hat{\phi}_1 \left| \psi \right\rangle - \left\langle \psi \right| \hat{\phi}_1^\dagger  \left| \psi \right\rangle=0,
\end{align} and therefore, the operator $\hat{\phi}_1$ representing one of the constraints for the ideal gas is hermitian. 

\section{Alternative choices for $\hat{\phi}_1$ in ideal gas model}\label{Appendix_B}

One may naturally ask what happens if we choose a different realization for the operator $\hat{\phi}_1$ instead of the symmetrized form used in Eq.~(\ref{phi1simetrico}). For instance, if we do not impose Hermiticity from the outset, a possible choice is
\begin{equation}
    \hat{\phi}_{1}=\hat{\pi}+\frac{\hat{q}\hat{p}}{k_{B}}. \label{outrophi1a}
\end{equation}
To achieve the corresponding physical states, we follow the same procedure as in Sec.~\ref{subsection_idealgas}: we impose the constraint conditions (\ref{condition}) simultaneously for $\hat{\phi}_1$ and $\hat{\phi}_2$, while realizing the canonical operators according to Eqs.~(\ref{q})–(\ref{pi}).

Since $\hat{\phi}_2$ remains unchanged, it still constrains the thermodynamic wave function to have the general form
\begin{equation}
\psi\left(\tau,q\right)=g\left(\tau\right)\exp\left[\frac{i}{\bbar}u(\tau, q)\right]. \label{psiDENOVO}
\end{equation}
The difference now lies in the constraint imposed by $\hat{\phi}_1$. It gives
\begin{equation}
    \hat{\phi}_{1}(\psi) =\left(\hat{\pi}+\frac{\hat{q}\hat{p}}{k_{B}}\right)\psi\left(\tau,q\right)=0, 
\end{equation}
which, when combined with the expression (\ref{psiDENOVO}), yields 
\begin{equation}
    \frac{dg\left(\tau\right)}{d\tau}=0. \label{g da wave}
\end{equation}
This condition fixes $g(\tau)$ as a constant, so that the final form of the thermodynamic wave function for the ideal gas (under the choice (\ref{outrophi1a}) for $\hat{\phi}_1$) is
\begin{equation}
    \psi\left(\tau,q\right)=\tilde{\alpha}\exp\left[\frac{i}{\bbar}u\right], \quad \tilde{\alpha} \in \mathbb{C}. \label{exemploNOsime}
\end{equation}
In this case, the temperature operator becomes naturally Hermitian, since
\begin{equation}
    \hat{\pi}\left(\psi\right) = \frac{\partial u}{\partial \tau}\psi (\tau,q).
\end{equation}
The same reasoning holds for the pressure operator.

These results differ from the case in which we reverse the operator ordering in the constraint, replacing $qp$ by $pq$. In that case, we have
\begin{align}
    \hat{\phi}_{1}(\psi) =\left(\hat{\pi}+\frac{\hat{p}\hat{q}}{k_{B}}\right)\psi\left(\tau,q\right)=\left(\hat{\pi}+\frac{-i\bbar+\hat{q}\hat{p}}{k_{B}}\right)\psi\left(\tau,q\right),
\end{align}
where the last equality follows directly from the commutation relation (\ref{commutadors ideal gas}). This modification changes the differential equation (\ref{g da wave}) to
\begin{equation}
    \frac{dg\left(\tau\right)}{d\tau}+\frac{g(\tau)}{k_{B}}=0,
\end{equation}
whose solution is 
\begin{equation}
g(\tau) = \bar{\alpha} e^{-\frac{\tau}{k_{B}}}, \quad \bar{\alpha}\in\mathbb{C}.    
\end{equation}
The corresponding thermodynamic wave function is then
\begin{equation}
    \psi\left(\tau,q\right)=\bar{\alpha} e^{-\frac{\tau}{k_{B}}}\exp\left[\frac{i}{\bbar}u(\tau,q)\right].
\end{equation}
This expression differs from the wave function (\ref{psi gas ideal}) obtained in the main text only by a factor of one-half in the real exponential term. Nevertheless, this case represents the least favorable scenario, as both $\hat{\phi}_1$ and $\hat{\pi}$ are non-Hermitian. As discussed in Sec.~\ref{subsec_pseudo}, however, all these operator realizations are physically equivalent within the pseudo-Hermitian framework, since they can be mapped into one another through suitable similarity transformations that preserve the underlying thermodynamic structure. 

\section{Pseudo-Hermitian systems}\label{Appendix_C}

\subsection{Pseudo-Hermiticity}

In standard quantum mechanics, the Hamiltonian $\hat{H}$ is required to be Hermitian ($\hat{H} = \hat{H}^\dagger$), which guarantees real energy eigenvalues and unitary time evolution, thus conserving probability. Pseudo-Hermitian quantum mechanics offers a generalization of this framework \cite{mostafazadeh2010}. A Hamiltonian $\hat{H}$ is called pseudo-Hermitian if it is not Hermitian ($\hat{H} \neq \hat{H}^\dagger$) but instead satisfies the relation
\begin{equation}
\hat{H}^\dagger = \hat{\Theta} \hat{H} \hat{\Theta}^{-1},
\end{equation}
where $\hat{\Theta}$ is a Hermitian, invertible operator known as a metric operator. If the metric $\hat{\Theta}$ is also positive-definite ($\hat{\Theta} > 0$), $\hat{H}$ is called quasi-Hermitian \cite{scholtz1992}. This positivity is crucial, as it allows for the re-definition of the inner product of the Hilbert space. A new inner product can be defined as:
\begin{equation}
\langle \phi , \psi \rangle_\Theta := \langle \phi , \hat{\Theta}  \psi \rangle .
\end{equation}
With respect to this new inner product, the pseudo-Hermitian $\hat{H}$ behaves \emph{as if} it were Hermitian. This ensures that its spectrum is real and that time evolution is unitary \emph{with respect to the new inner product}, $\langle \psi(t), \psi(t) \rangle_\Theta = 1$.

\subsection{The Time-Dependent Formalism}

Introducing explicit time-dependence, $\hat{H} = \hat{H}(t)$,  modifies the formalism significantly. The central challenge is ensuring that the time evolution remains unitary. For the probability defined by the $\hat{\Theta}$-inner product to be conserved, the metric operator itself must, in general, also become time-dependent, i.e., $\hat{\Theta} = \hat{\Theta}(t)$. The time evolution of a state $|\psi(t)\rangle$ is governed by the standard Schrödinger equation
\begin{equation}
i\hbar \frac{\partial}{\partial t} |\psi(t)\rangle = \hat{H}(t) |\psi(t)\rangle.
\end{equation}
However, for the $\hat{\Theta}$-norm $\langle \psi(t), \psi(t) \rangle_{\Theta}$ to be conserved, the metric and the Hamiltonian must satisfy a dynamical relation derived from $\frac{d}{dt} \langle \psi(t) | \hat{\Theta}(t) | \psi(t) \rangle = 0$. This leads to the constraint
\begin{equation}
i\hbar \frac{\partial \hat{\Theta}(t)}{\partial t} = \hat{H}^\dagger(t) \hat{\Theta}(t) - \hat{\Theta}(t) \hat{H}(t). \label{eq:app_constraint}
\end{equation}
This equation is known as time-dependent quasi-Hermitian relation, being the key condition for ensuring unitary time evolution in a time-dependent pseudo-Hermitian system.

The formalism is often implemented using a time-dependent Dyson map, $\hat{\eta}(t)$, which relates the non-Hermitian Hamiltonian $\hat{H}(t)$ to an equivalent, time-dependent \textit{Hermitian} Hamiltonian $\hat{h}(t)$ \cite{fring2016}. This map defines a new state $|\chi(t)\rangle$
\begin{equation}
|\chi(t)\rangle = \hat{\eta}(t) |\psi(t)\rangle.
\end{equation}
The evolution of this new state is governed by a conventional Hermitian Hamiltonian $\hat{h}(t) = \hat{h}^\dagger(t)$
\begin{equation}
i\hbar \frac{\partial}{\partial t} |\chi(t)\rangle = \hat{h}(t) |\chi(t)\rangle.
\end{equation}
By differentiating the definition of $|\chi(t)\rangle$ and substituting the two Schrödinger equations, we find the relationship between the Hamiltonians
\begin{equation}
\hat{h}(t) = \hat{\eta}(t) \hat{H}(t) \hat{\eta}^{-1}(t) + i\hbar \bigg(\frac{\partial \hat{\eta}(t)}{\partial t}\bigg) \hat{\eta}^{-1}(t).
\end{equation}
This transformation maps the complex dynamics of $\hat{H}(t)$ to the familiar, unitary dynamics of $\hat{h}(t)$. The metric operator is given by $\hat{\Theta}(t) = \hat{\eta}^\dagger(t) \hat{\eta}(t)$, which can be shown to satisfy the constraint in Eq. \eqref{eq:app_constraint} automatically.

\subsection{Observables vs. Generators of Evolution}

In this formalism, physical observables are represented by different operators in the two representations. In the Hermitian representation ($\chi$-space), an observable is a standard Hermitian operator $\hat{o} = \hat{o}^\dagger$. In the pseudo-Hermitian representation ($\psi$-space), the corresponding observable $\hat{O}$ is related by the similarity transformation
    \begin{equation}
    \hat{O}(t) = \hat{\eta}^{-1}(t) \hat{o}(t) \hat{\eta}(t).
    \end{equation}
This transformation ensures that the physical expectation value is identical in both representations. Starting from the Hermitian representation:
\begin{align}
\label{eq:app_expectationvalue}
\langle \hat{o} \rangle_{\chi} 
= \frac{\langle \chi , \hat{o}  \chi \rangle}{\langle \chi ,   \chi \rangle} 
= \frac{\langle \hat{\eta} \psi, \hat{o} \hat{\eta} \psi \rangle}{\langle \hat{\eta} \psi, \hat{\eta} \psi \rangle} 
= \frac{\langle \psi, \hat{\eta}^\dagger \hat{o} \hat{\eta} \psi \rangle}{\langle \psi, \hat{\eta}^\dagger \hat{\eta} \psi \rangle} 
= \frac{\langle \psi , (\hat{\eta}^\dagger \hat{\eta}) (\hat{\eta}^{-1} \hat{o} \hat{\eta}) \psi \rangle}{\langle \psi, \hat{\eta}^\dagger \hat{\eta} \psi \rangle} 
= \frac{\langle \psi , \hat{\Theta}(t) \hat{O}  \psi \rangle}{\langle \psi , \hat{\Theta}(t) \psi \rangle} 
= \frac{\langle \psi , \hat{O} \psi \rangle_{\Theta(t)}}{\langle \psi , \psi \rangle_{\Theta(t)}}
= \langle \hat{O} \rangle_{\psi}.
\end{align}
This calculation also proves that any observable $\hat{O}$ in the $\psi$-space must be quasi-Hermitian, satisfying the condition $\hat{O}^\dagger(t) = \hat{\Theta}(t) \hat{O}(t) \hat{\Theta}^{-1}(t)$, which guarantees its expectation values are real.

A crucial consequence of this formalism is that the non-Hermitian Hamiltonian $\hat{H}(t)$ is \emph{not an observable}. If we check this quasi-Hermiticity condition for $\hat{H}(t)$, we find it fails. The dynamical constraint (Eq. \eqref{eq:app_constraint}) shows:
\begin{equation}
\hat{H}^\dagger(t) = \hat{\Theta}(t) \hat{H}(t) \hat{\Theta}^{-1}(t) - i\hbar \bigg(\frac{\partial \hat{\Theta}(t)}{\partial t}\bigg) \hat{\Theta}^{-1}(t).
\end{equation}
Because $\partial_t \hat{\Theta}(t) \neq 0$ for the general time-dependent case, $\hat{H}^{\dagger}(t) \neq \hat{\Theta}(t) \hat{H}(t) \hat{\Theta}^{-1}(t)$. Therefore, $\hat{H}(t)$ is interpreted only as the \emph{generator of time evolution}, not as the observable for the system's energy. The physical, measurable energy corresponds to the associated Hermitian Hamiltonian $\hat{h}(t)$, which \textit{is} an observable by definition.

\end{document}